\documentclass[reprint,aps,prl,superscriptaddress,english,nolongbibliography]{revtex4-2}

\usepackage{graphicx}
\usepackage[colorlinks=true,citecolor=blue,urlcolor=blue,linkcolor=blue]{hyperref}
\usepackage{amsmath,amssymb,bm,mathrsfs,amsfonts}

\usepackage{color,colortbl,hhline}
\usepackage[table,xcdraw,dvipsnames]{xcolor}
\usepackage[caption=false,subrefformat=parens,labelformat=empty]{subfig}

\usepackage[normalem]{ulem}

\newcommand{\sbra}[1]{\langle #1 |}
\newcommand{\sket}[1]{| #1 \rangle}
\newcommand{\bra}[1]{\left\langle #1 \right|}
\newcommand{\ket}[1]{\left| #1 \right\rangle}

\definecolor{mygray}{RGB}{153, 153, 153}
\newcommand{\gsquare}[0]{\textcolor{mygray}{\blacksquare}}

\usepackage{tikz}

\newcommand{\urmove}[0]{%
\begin{tikzpicture}[baseline=0.1ex]%
\draw[thick, red] (0,1.5ex) -- (1.5ex,1.5ex);%
\draw[thick, red] (1.5ex,1.5ex) -- (1.5ex,0);%
\end{tikzpicture}%
}

\newcommand{\ulmove}[0]{%
\begin{tikzpicture}[baseline=0.1ex]%
\draw[thick, red] (0,1.5ex) -- (1.5ex,1.5ex);%
\draw[thick, red] (0,1.5ex) -- (0,0);%
\end{tikzpicture}%
}

\newcommand{\dlmove}[0]{%
\begin{tikzpicture}[baseline=0.1ex]%
\draw[thick, red] (0,1.5ex) -- (0,0);%
\draw[thick, red] (0,0) -- (1.5ex,0);%
\end{tikzpicture}%
}

\newcommand{\drmove}[0]{%
\begin{tikzpicture}[baseline=0.1ex]%
\draw[thick, red] (0,0) -- (1.5ex,0);%
\draw[thick, red] (1.5ex,1.5ex) -- (1.5ex,0ex);%
\end{tikzpicture}%
}

\newcommand{\vvmove}[0]{%
\begin{tikzpicture}[baseline=0.1ex]%
\draw[thick, red] (0,0) -- (0,1.5ex);%
\draw[thick, red] (1.5ex,0) -- (1.5ex,1.5ex);%
\end{tikzpicture}%
}

\newcommand{\hhmove}[0]{%
\begin{tikzpicture}[baseline=0.1ex]%
\draw[thick, red] (0,0) -- (1.5ex,0);%
\draw[thick, red] (1.5ex,1.5ex) -- (0,1.5ex);%
\end{tikzpicture}%
}

\begin{document}

\title{Localization and melting of interfaces in the two-dimensional quantum Ising model}

\author{Federico Balducci}
\affiliation{SISSA -- via Bonomea 265, 34136, Trieste, Italy}
\affiliation{The Abdus Salam ICTP -- Strada Costiera 11, 34151, Trieste, Italy}
\affiliation{INFN Sezione di Trieste -- Via Valerio 2, 34127 Trieste, Italy}
\author{Andrea Gambassi}
\affiliation{SISSA -- via Bonomea 265, 34136, Trieste, Italy}
\affiliation{INFN Sezione di Trieste -- Via Valerio 2, 34127 Trieste, Italy}
\author{Alessio Lerose}
\affiliation{Department of Theoretical Physics, University of Geneva -- Quai Ernest-Ansermet 30, 1205 Geneva, Switzerland}
\author{Antonello Scardicchio}
\affiliation{The Abdus Salam ICTP -- Strada Costiera 11, 34151, Trieste, Italy}
\affiliation{INFN Sezione di Trieste -- Via Valerio 2, 34127 Trieste, Italy}
\author{Carlo Vanoni}
\email{cvanoni@sissa.it}
\affiliation{SISSA -- via Bonomea 265, 34136, Trieste, Italy}
\affiliation{The Abdus Salam ICTP -- Strada Costiera 11, 34151, Trieste, Italy}
\affiliation{INFN Sezione di Trieste -- Via Valerio 2, 34127 Trieste, Italy}

\date{\today}

\begin{abstract}
    We study the non-equilibrium evolution of coexisting ferromagnetic domains in the two-dimensional quantum Ising model---a setup relevant in several contexts, from quantum nucleation dynamics and false-vacuum decay scenarios to recent experiments with Rydberg-atom arrays. We demonstrate that the quantum-fluctuating interface delimiting a large bubble can be studied as an effective one-dimensional system through a ``holographic'' mapping. For the considered model, the emergent interface excitations map to an integrable chain of fermionic particles. We discuss how this integrability is broken by geometric features of the bubbles and by corrections in inverse powers of the ferromagnetic coupling, and provide a lower bound to the timescale after which the bubble is ultimately expected to melt. Remarkably, we demonstrate that a symmetry-breaking longitudinal field gives rise to a robust ergodicity breaking in two dimensions, a phenomenon underpinned by Stark many-body localization of the emergent fermionic excitations of the interface.
\end{abstract}

\maketitle

Strongly interacting quantum many-body systems evolving out of equilibrium are generically expected to locally relax to  thermodynamic equilibrium after a short transient~\cite{Rigol2008Thermalization,Polkovnikov2011Colloquium}. In several cases, however, microscopic interactions support long-lived dynamical stages away from equilibrium. These anomalous nonequilibrium states attract much interest, as they facilitate the realization of unconventional phases of matter. While quenched disorder gives rise to the strongest form of ergodicity breaking~\cite{Anderson1958Absence,Basko2006MetalInsulator,Abanin2019Colloquium}, characterized by emergent integrals of motion \cite{serbyn2013local,huse2014phenomenology,ros2015integrals,imbrie2017local}, suppression of thermalization may arise from a variety of mechanisms in translationally invariant Hamiltonian systems, including configurational disorder~\cite{DeRoeck2014Asymptotic,DeRoeck2014Scenario,Carleo2012Localization,Michailidis2018Slow,Yao2016Quasi,Papic2015Many,Schiulaz2015Dynamics,Smith2017Disorder,Brenes2018Many,Karpov2021Disorder,Hart21logarithmic}, kinetic constraints \cite{Surace2020Lattice,Pancotti2020Quantum,Sierant2021Constraints,Oppong2020Probing,Orito21nonthermalized}, long-range interactions~\cite{KastnerPRL11_Diverging,mori2018prethermalization,neyenhuis2017observation,LerosePappalardiPRR20,Defenu2021Metastability,LeroseDWLR,GorshkovConfinement}, confinement of elementary excitations~\cite{Kormos2017Real,Robinson2019Nonthermal,Robinson2019Signatures,Lin2017Quasiparticle,Mazza2019Suppression,Lerose2020Quasilocalized,Verdel2020Real,Pai2020Fractons,Chanda2020Confinement,Magnifico2020Real,birnkammer2022prethermalization,Park19glassy}, Hilbert-space fragmentation~\cite{Nandkishore2019Fractons,Pretko2020Fracton,Sala2020Ergodicity,Khemani2020Localization,bastianello2021fragmentation}, or quantum many-body scars~\cite{Turner2018Weak,Serbyn2021Quantum}.
These phenomena are expected to be fragile to generic perturbations, which should cause a slow drift toward eventual thermalization.

Much insight on non-ergodic behavior is gained from one-dimensional (1$d$) systems, for which advanced analytical~\cite{korepin1993quantum,giamarchi2003quantum} and numerical~\cite{Schollwock2011Density} techniques are available. The nonequilibrium and possibly non-ergodic evolution of higher-dimensional quantum systems, instead, is a largely uncharted territory. While the development of theoretical tools to analyze their dynamics  stands as a formidable  challenge, recent experimental advances allow an unprecedented degree of engineering and control of two-dimensional (2$d$) arrays of two-level systems~\cite{labuhn2016tunable,Guardado2018Probing,scholl2021quantum,ebadi2021quantum,Bluvstein2022Quantum}, exhibiting forms of ergodicity breaking~\cite{bluvstein2021controlling}. Moreover, it was recently suggested that pseudorandom disorder may stabilize many-body localization in 2$d$ systems~\cite{agrawal2022note,Strkalj2022Coexistence,crowley2022mean}.

\begin{figure}[t]
    \centering
    \includegraphics[width=\columnwidth]{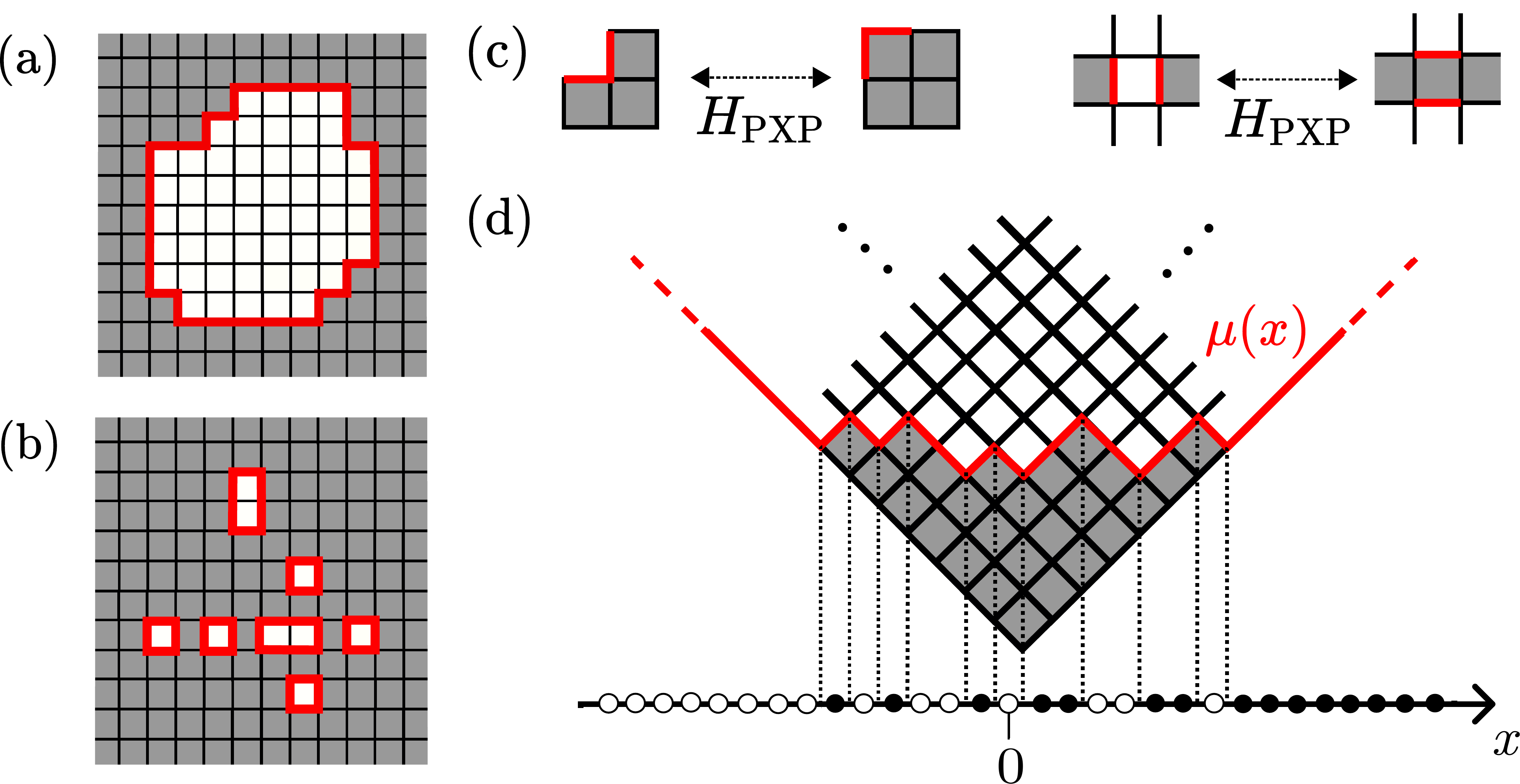}
    \caption{(a) Example of a convex bubble of ``down'' spins ($\Box$) in a sea of ``up'' spins ($\gsquare$) (see main text for details).
    (b) One of the maximally split configurations of the bubble in panel (a).
    (c) Schematic representation of the allowed transitions in $H_{\mathrm{PXP}}$preserving the total domain wall length. 
    (Other possible orientations are not displayed.)
    (d) Mapping of an interface corner onto a fermionic chain.}
    \label{fig:PXPHam}
\end{figure}

In this work, we formulate a novel general approach to analyze the unitary dynamics of 2$d$ quantum  lattice systems with multiple degenerate or nearly-degenerate vacuum states, originating from a broken discrete symmetry. 
Our approach allows us to ascertain the emergence of non-ergodic  dynamical regimes,  rooted in the spatial coexistence of multiple regions occupied by different vacua separated by ``smooth'' interfaces.
For concreteness, we consider the quantum Ising model in a weak external magnetic field, and inspect the non-equilibrium evolution of large domains (or bubbles) of negatively magnetized spins initially prepared in a background of positively magnetized ones, as in Fig.~\hyperref[fig:PXPHam]{\ref{fig:PXPHam}a}. 
The thermalization of these atypical highly excited states may be hindered by emergent kinetic constraints or high energetic barriers. We show that one can successfully describe the \emph{prethermal} evolution of such a domain by reducing it to the motion of its quantum-fluctuating interface, which makes the problem effectively 1$d$. We study the resulting dynamics using complementary analytical and numerical tools, which allow us to unveil surprising non-ergodic features and characterize their robustness. In particular, we find that the dynamics of the quantum interface map exactly onto those of a chain of fermions in an external field, undergoing Wannier-Stark (many-body) localization for an infinite (large but finite) Ising coupling~\cite{Refael2019From,Schulz2019Stark,scherg2021observing,Morong2021Observation}.
This generalizes the quasilocalization of domain walls in the 1$d$ setting~\cite{Mazza2019Suppression,Lerose2020Quasilocalized}.

The phenomena investigated here have profound connections with relevant issues in different contexts. In fact, they are the quantum counterpart of well-studied phenomena in classical stochastic dynamics, e.g.\ the fluctuations of interfaces between coexisting phases \cite{Krapivsky2004Dynamics,Krapivsky2012Limiting} or---when the symmetry is explicitly broken---the nucleation of ``true-vacuum'' bubbles in a ``false-vacuum'' background~\cite{bray1994theory,onuki2002phase}. Similarly, our findings are relevant to understanding Coleman's false-vacuum decay~\cite{Coleman1977False,coleman1988aspects,Rutkevich1999Decay,Lagnese2021False} in lattice systems, specifically the route from highly non-equilibrium states with sparse true-vacuum bubbles to the eventual thermal state. In both these settings, bubbles delimited by ``smooth'' interfaces emerge as natural states. Moreover, our mapping from $2d$ Ising to $1d$ confined fermions can also be interpreted as a toy model of duality between a theory of string (the domain wall) in 2+1 dimensions, and a theory of particles in one less dimension~\cite{maldacena1999large,gubser1998gauge,witten1998anti}, which becomes integrable in the limit of infinite string tension; in our case, the confinement of the fermions is not due to their interaction but to an external potential. Finally, the dynamics at the corners of large bubbles in a $2d$ quantum ferromagnet turns out to be related to a measure concentration phenomenon for random Young diagrams, a well-known result to the mathematical community \cite{Logan1977Variational,Vershik1977AsymPlancherel,Vershik1977AsymMaximal,Okounkov2000Random,Krapivsky2021Stochastic}.

\emph{Model.~---}
We consider the dynamics of the quantum Ising model on a 2$d$ square lattice, with Hamiltonian
\begin{equation}
    \label{eq:Ising2d_ham}
    H_{\mathrm{Is}} = -J\sum_{\langle i,j\rangle} \sigma_i^z \sigma_j^z - g \sum_i \sigma_i^x - h \sum_i \sigma_i^z.
\end{equation}
Here $\sigma_i^{x,y,z}$ are Pauli matrices at site $i\in\mathbb{Z}^2$, $g>0$ and $h$ are the transverse and longitudinal magnetic fields, respectively,  and $J>0$ is the ferromagnetic coupling.

In this work, we assume $J\gg |h|, g$, which allows us to treat $g$ and $h$ as weak perturbations. Correspondingly, the dynamics \emph{quasi-conserves} the domain-wall total length $\ell$: considering states in the unperturbed $\sigma^z$-basis, the transitions in $H_{\mathrm{Is}}$  corresponding to spin flips (with amplitude $g$) that change $\ell \to \ell \pm 2$ or $\pm 4$, encounter large energy mismatch $\pm 4J$ or $\pm 8J$, and thus can be adiabatically eliminated by a Schrieffer-Wolff transformation~\cite{FrohlichSchriefferWolff,LossSchriefferWolff}. A direct application of rigorous prethermalization bounds~\cite{Abanin2017Rigorous} shows that the dressed domain-wall length operator $D=\sum_{\langle i,j \rangle} (1 - \sigma^z_i \sigma^z_j )/2$ is approximately conserved for a long time $T_{\mathrm{preth}} \gtrsim (C/g) \exp[ c J/\max(g,|h|)]$ (where $c$ and $C$ are numerical constants). More specifically, the evolution of local observables from their values at time $t=0$ is well approximated for $t \le T_{\mathrm{preth}}$ by 
that governed by an effective Hamiltonian $H_{\mathrm{eff}}$, constructed order by order in $J^{-1}$, which conserves $D$: $[H_{\mathrm{eff}},D]=0$. Thus, $H_{\mathrm{eff}}$ preserves the Hilbert space sectors identified by $\ell$: $\mathcal{H}=\bigoplus_\ell \mathcal{H}_\ell$.

We first consider the limit $J=\infty$. The lowest-order $H_{\mathrm{eff}}$ arises from projecting Eq.~\eqref{eq:Ising2d_ham} onto each sector $\mathcal{H}_\ell$, yielding a constrained Hamiltonian of ``PXP'' type~\cite{Fendley2004Competing,Yoshinaga2021Emergence}, closely related to that emerging in Rydberg-blockaded atomic arrays~\cite{Jaksch200Fast,Lukin2001Dipole,Bernien2017Probing}:
\begin{equation}
    \label{eq:PXPEff}
    H_{\mathrm{PXP}} = -  \sum_i h \sigma_i^z  
    + g  \Big( \sket{\ulmove}_i{}_i\sbra{\drmove} + \sket{\dlmove}_i{}_i\sbra{\urmove} + \sket{\hhmove}_i{}_i\sbra{\vvmove} + \mathrm{H.c.} \Big).
\end{equation}
The notation refers to local changes in domain wall configurations caused by the allowed spin flips, see Fig.~\hyperref[fig:PXPHam]{\ref{fig:PXPHam}c}.

\emph{Fragmentation and bubbles.~---} The dynamics of $H_{\mathrm{PXP}}$ in Eq.~\eqref{eq:PXPEff} within each sector $\mathcal{H}_\ell$ is strongly affected by the constraints. In fact, it is easy to see that each $\mathcal{H}_\ell$ fractures into exponentially many (in $\ell$) disconnected subsectors, as also noted in Ref.~\cite{Yoshinaga2021Emergence}. The simplest example is given by an isolated spin surrounded by a thick belt of opposite spins; more generally, any bubble of spins with reversed polarization compared to that of a sufficiently extended surrounding region can never expand beyond the smallest rectangle fully containing the initial bubble. Accordingly, each bubble forms an isolated dynamical system and therefore below we will focus on a subsector with a single bubble of perimeter $\ell$ \footnote{Note that also for finite but large $J<\infty$, elementary perturbation theory arguments show that the mobility of a bubble as a whole is exponentially suppressed with its size. For all accessible times, one is thus only concerned with the ``internal'' bubble dynamics analyzed below, rather than with its very slow global delocalization in space.}.  

In order to investigate ergodicity, it is convenient to consider a very atypical initial bubble, delimited by an extended ``smooth'' convex interface~\footnote{Note that the notion of convexity employed here is a lattice generalization of the conventional notion: a connected subset of a lattice is said to be convex whenever any segment parallel to the lattice axes joining two points of the subset belongs to the subset.}, see e.g.\ Fig.~\hyperref[fig:PXPHam]{\ref{fig:PXPHam}a}. The evolution ruled by $H_{\mathrm{PXP}}$ can be pictured as a quantum walk on the graph of the exponentially many possible configurations of the domain wall of length~$\ell$. At long times, the domain wall may be naively expected to explore all the dynamically accessible configurations and, in particular for $h=0$, to dissolve into $\mathcal{O}(\ell)$ small bubbles---the most entropic macrostate (see e.g.\ Fig.~\hyperref[fig:PXPHam]{\ref{fig:PXPHam}b}). At short times, however, the dynamics generates local quantum fluctuations of the interface starting from its \emph{corners}, as neither inner/outer spins, nor spins adjacent to a flat portion of the interface can flip. Thus, we first consider the evolution of an isolated corner.

\emph{Isolated corner.~---} Let us consider a right-angled corner with infinitely long sides, ``down'' spins inside and ``up'' outside, as in Fig.~\hyperref[fig:PXPHam]{\ref{fig:PXPHam}d}. The evolution governed by $H_{\mathrm{PXP}}$ may eventually flip inner spins starting from the one at the apex, via transitions of type $\sket{\ulmove}\sbra{\drmove}$ and $\sket{\dlmove}\sbra{\urmove}$ and their conjugates. Crucially, interface-splitting terms such as $\sket{\hhmove}\sbra{\vvmove}$ are inconsequential in this evolution: this fact allows for the \emph{exact} mapping of the problem onto a fictitious $1d$ fermionic chain as follows.

First, we note that the configurations of the domain wall originating from the isolated corner are in one-to-one correspondence with \emph{Young diagrams}, i.e.\ the possible configurations of a collection of boxes, arranged in left-justified rows, and stacked in non-increasing order of length. Figure~\hyperref[fig:PXPHam]{\ref{fig:PXPHam}d} reports an example, where the domain wall is the red line and the corresponding (rotated) Young diagram is highlighted in gray. Second, a $1d$ fermionic representation is obtained by means of a ``holographic'' projection \cite{Okounkov2001Infinite,Okounkov2006Quantum,Dijkgraaf2009Quantum,araujo2021quantum,Krapivsky2004Dynamics,Krapivsky2012Limiting,Krapivsky2021Stochastic}, illustrated again in Fig.~\hyperref[fig:PXPHam]{\ref{fig:PXPHam}d}. Considering the Young diagram rotated by $\pi/4$, we project its segments onto a horizontal chain labelled by $x\in\mathbb{Z}$, with the prescription that, following the interface from left to right, we associate to each up-going line a site occupied by a particle and to each down-going line an empty site. This construction produces a representation of the accessible Hilbert space as that of a $1d$ fermionic chain, described by annihilation/creation operators $\psi_x$/$\psi_x^{\dagger}$ such that $\{\psi_x, \psi^{\dagger}_y \} = \delta_{xy}$. Based on this mapping, one gets convinced that the resulting fermionic Hamiltonian reads~\cite{SupplMat}
\begin{equation}
    \label{eq:hamFerm}
    H_{\mathrm{F}} = - g\sum_{x\in\mathbb{Z}} \left( \psi_x^{\dagger} \psi_{x+1} + \mathrm{h.c.} \right) + 2 h \sum_{x\in\mathbb{Z}} x \, \psi_x^{\dagger} \psi_x^{\phantom{\dagger}},
\end{equation} 
since a fermion hop corresponds to an elementary domain-wall move of the form $\sket{\ulmove}{}\sbra{\drmove}$, and the last sum counts how many squares have been flipped inside the 2$d$ corner, up to a constant. The full corner is represented by a ``voltage-bias'' Fermi sea, i.e.\ a $1d$ domain-wall state: $\ket{\Psi_0} \equiv \big(\prod_{x>0}\psi_x^{\dagger}\big) \ket{\emptyset}$, where $\ket{\emptyset}$ is the vacuum state. 

Remarkably, $H_{\mathrm{F}}$ in Eq.~\eqref{eq:hamFerm} describes  \emph{non-interacting} fermions hopping along the chain with amplitude $g$, and subject to a constant field of strength $2h$, i.e.~a Wannier-Stark ladder~\cite{grosso2013solid}. The fact that the emergent excitations are non-interacting allows us to compute the evolution of the isolated corner for arbitrary values of $g$ and $h$ using exact and asymptotic methods. The observable of primary interest is the shape of the interface. From the construction of the mapping above, the height operator $\mu(x)$ of the interface---which measures its distance along the vertical direction from the horizontal straight line (c.f.\ Fig.~\hyperref[fig:PXPHam]{\ref{fig:PXPHam}d})---is related to the fermion density profile as $ \mu(x)=\sum_{y\le x} (2 \psi_y^{\dagger} \psi_y^{\phantom{\dagger}}-1) + \text{const}$; we choose the constant such that $\langle \mu(0) \rangle \equiv \langle \Psi_0 | \mu(0) |\Psi_0 \rangle = 0$. Thus, we extract the time-evolving shape of the interface by computing the average fermion density $\langle \psi_x^{\dagger}(t) \psi_x^{\phantom{\dagger}}(t) \rangle $.

The Hamiltonian~\eqref{eq:hamFerm} is diagonalized by the eigenstates $\phi_x^{(y)} = J_{x-y}(g/h)$ with eigenvalues $E^{(y)}=2hy$, where $y\in\mathbb{Z}$ and $J_{\nu}(z)$ is the Bessel function of the first kind. The spectrum forms an equispaced ladder, with spatially localized eigenstates related to each other via rigid translations. Thus, the evolution from an arbitrary configuration of the particles will exhibit persistent coherent oscillations of the density profile, with frequency $|h|/\pi$, interpreted as  Bloch oscillations of each particle over $\sim g/|h|$ lattice sites in the dc external field. For $h\to0$, this Wannier-Stark localization leaves room to delocalized plane-wave eigenstates and ballistic evolution of the fermion density. 

These results find immediate applications in the original $2d$ problem: arbitrary corner-like domain walls deform periodically in time and locally in space, preserving memory of the initial shape for arbitrarily long times. As $h$ decreases, these oscillations become slower and extend over a longer localization length. Ultimately, for $h\to0$, the corner is indefinitely eroded at constant speed.

\begin{figure}
    \centering
    \includegraphics[width=\columnwidth]{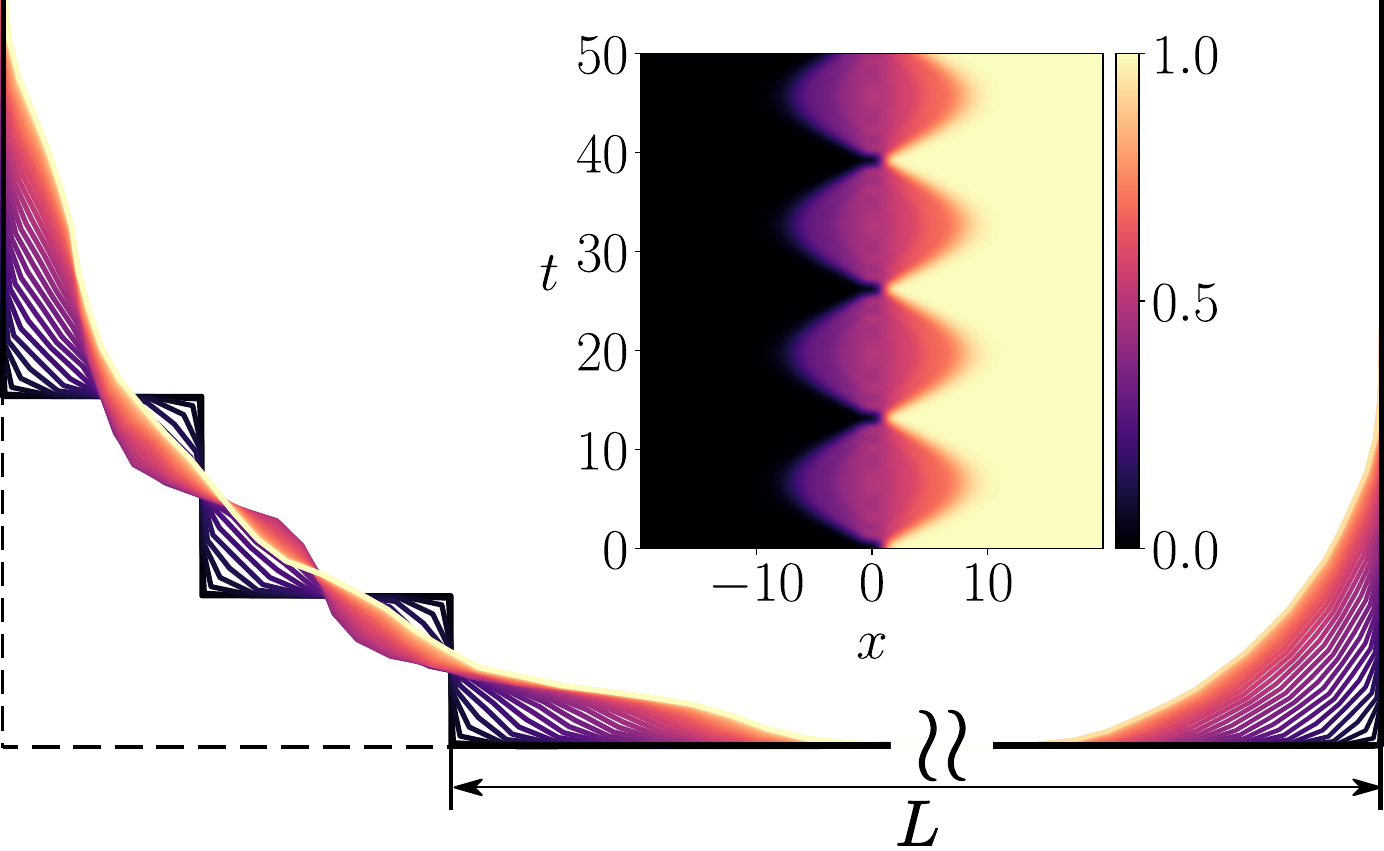}
    \caption{Snapshots of the time evolution of two corners separated by a large distance $L$, for $g=1$, $h=0.24$. The black initial profile evolves periodically as indicated by the various colors, reaching the yellow line at half period $t = \pi/(2|h|)$ before receding towards the initial condition.
    (Inset) Time evolution of the density profile $\langle \psi^\dagger_x(t)\psi_x(t)\rangle$ of the fermionic chain, starting from the state $|\Psi_0 \rangle$ corresponding to the right corner of the main figure. }
    \label{fig:doubleCorner}
\end{figure}

The large-scale behavior of the interface can be captured by a suitable continuum limit of the exact solution of the dynamics (well known in the context of free fermions~\cite{Antal1999Transport,SupplMat}), which yields
\begin{equation}
    \label{eq:contLimit}
    \langle\mu(x,t)\rangle =  \frac{2g\,|\sin(ht)|}{|h|} \, \Omega \left(\frac{h\, x}{2g \sin(ht)}\right),
\end{equation}
where $\Omega$ is the same function defining the asymptotic shape of large, random Young diagrams~\cite{Logan1977Variational,Vershik1977AsymPlancherel,Vershik1977AsymMaximal,Okounkov2000Random}:
\begin{equation}
\label{eq:Okounkov}
    \Omega(v)=
    \begin{cases}
        |v| &\mbox{for}\quad  |v| \geq 1, \\
        \frac{2}{\pi} \left(\sqrt{1-v^2} + v \arcsin v \right) &\mbox{for}\quad  |v| < 1.
\end{cases}
\end{equation}
For $h\to 0$, Eq.~\eqref{eq:contLimit} predicts $\langle \mu(x,t) \rangle_{h=0} = 2gt \, \Omega( x/2gt)$, i.e.\ the anticipated ballistic melting of the corner in the absence of the longitudinal field. In the right corner of the main panel of Fig.~\ref{fig:doubleCorner}, we report the exact interface evolution expressed by Eq.~\eqref{eq:contLimit}; the inset shows the corresponding evolution of the fermionic density operator.

The evolution of an arbitrary initial classical configuration connected with the infinite corner, e.g.~the one given by the black line in the left part of Fig.~\ref{fig:doubleCorner}, can be analyzed exactly by using the same mapping to the fermionic chain. Figure~\ref{fig:doubleCorner} reports the time-evolved domain boundary over a half-period $0<t<\pi/(2|h|)$, as the subsequent evolution brings it back to its initial configuration (black line), periodically in time. Memory of the initial shape is retained for all times---a $2d$ manifestation of the underlying Wannier-Stark localization of the emergent interface degrees of freedom via our ``holographic'' mapping.

\emph{Generic ``smooth'' bubbles.~---} After solving the dynamics of isolated corners we can address the one of a large $2d$ bubble. For a generic initial shape at fixed perimeter the evolution can be extremely complicated, due to the fact that the interface-splitting terms, $\sket{\hhmove}\sbra{\vvmove}$ and $\sket{\vvmove}\sbra{\hhmove}$, are as important as (or even more important than) the interface-hopping terms of the form $\sket{\ulmove}\sbra{\drmove}$, $\sket{\dlmove}\sbra{\urmove}$ and conjugates. This generically prevents the very possibility of mapping the $2d$ dynamics onto an effective $1d$ model. As anticipated above, such a mapping is only expected to be feasible when the interface delimiting a bubble is sufficiently ``smooth'', i.e.~when its (coarse-grained) local slope varies slowly in space and the abrupt variations due to extended corners are sufficiently dilute. Since we showed that corners remain spatially localized for $h\neq0$, the next natural issue consists in considering the presence of multiple corners along the interface, separated by a distance $L$, as shown in Fig.~\ref{fig:doubleCorner}.

Leaving the detailed description of a generalized mapping of the interface onto a $1d$ quantum system to future work, here we content ourselves with bounding from below the timescale over which consecutive localized corners separated by a flat portion of length $L$ of the interface start ``interacting''. This bound is based on estimating the probability $P_L$ of finding two fermionic particles, each coming from a isolated corner, halfway on the flat interface. Due to Wannier-Stark localization, $P_L$ is extremely suppressed when $L$ is larger than the localization length, i.e.\ $L\gg g/|h|$. For $L\gg 1$, using the exact solution of the dynamics, $P_L$ can be estimated as $P_L \sim J_{L/2}(2g/h)^2 \sim (2e g /L |h|)^{L} / (\pi L)$ \cite{Balducci2022Interface}. Consequently, a lower bound on the timescale below which the isolated evolution of individual corners is an accurate description of the global bubble dynamics is $T_{\mathrm{corner}}\sim \exp[L\ln L - L\ln (2e g/|h|)] / g$. This timescale diverges faster than exponentially upon increasing $L$. Accordingly, the interface of sufficiently ``smooth'' large bubbles remains nearly pinned for extremely long times, exhibiting a remarkable memory of their initial shape for \emph{arbitrary} ratios $0\le g/|h| < \infty$.

\emph{Beyond $J=\infty$.~---} 
The above analysis applies to $H_{\mathrm{PXP}}$ in Eq.~\eqref{eq:PXPEff}, which gives an accurate description of the prethermal dynamics over a timescale $T_{\mathrm{int}}\sim J/g^2$. In practice, when $J$ is large but finite, the lower bound $T_{\text{corner}}$ found above might exceed $T_{\mathrm{int}}$. This makes it compelling to investigate higher-order corrections in $1/J$ to $H_{\mathrm{PXP}}$. We argue that the second-order corrections break the integrability of the interface Hamiltonian derived from $H_{\mathrm{PXP}}$. In this case, we provide solid numerical evidence of a surprising lack of ergodicity of the $2d$ bubble dynamics persisting for finite $J$, which is rooted in the robustness of the Wannier-Stark localization of the interface's fermionic degrees of freedom to weak many-body interactions~\cite{Refael2019From,Schulz2019Stark}.

The $1/J$-corrections to $H_{\mathrm{PXP}}$ can be derived via a standard Schrieffer-Wolff transformation (see above). The resulting constrained Hamiltonian contains more complicated quasi-local transitions of the domain-wall configurations~\cite{Balducci2022Interface}. In terms of fermions, it 
corresponds to having $H^{\prime}_{\mathrm{F}} = H_{\mathrm{F}} + V_{\mathrm{F}}$, with 
\begin{equation}
\begin{split}
    &V_{\mathrm{F}} = -\frac{g^2}{4J} \sum_x \left(\psi^\dagger_x \psi_{x+2}+ \mathrm{h.c.}\right) \\
    &+ \frac{g^2}{4J}\sum_x\left(2\psi^\dagger_x \psi^\dagger_{x+1} \psi_{x+1}\psi_{x+2}+ \mathrm{h.c.}-3\psi^\dagger_{x} \psi_{x}\psi^\dagger_{x+1} \psi_{x+1}\right).
\end{split}
\label{eq:JCorrect}
\end{equation}
Two-body interactions in the second line break the emergent integrability of $H_{\mathrm{F}}$.

To develop an intuition of the effects of $V_F$, we start by arguing that, for sufficiently strong longitudinal field~$h$, the perturbed system remains non-ergodic. The argument is inspired by Ref.~\cite{Basko2006MetalInsulator}: in the integrable limit $J=\infty$ the eigenfunctions are localized, with localization length given by $\xi=1 / \mathrm{IPR}$, being $\mathrm{IPR}=\sum_k J_k^4(g / h) = 1 / \pi \int_0^{\pi} d\theta \, J_0^2( \sqrt{2-2\cos{\theta}} g/ h)$ a function of the ratio $g/h$. In a box of size $\xi$, the maximum energy difference between localized orbitals is $|h| \xi^2$, whereas the number of states is $2^{\xi}$. Interactions may be insufficient to restore ergodicity when their strength $\lambda \simeq g^2/J$ is smaller than the average local level spacing $\delta_{\xi}=|h| \xi^2/2^{\xi}$. Using the dimensionless ratios $x=g/|h|$ and $\varepsilon=g/|J|$, we find the heuristic criterion $\varepsilon < \xi^2(x)/(x 2^{\xi(x)})$, which is always satisfied for $0\le x \lesssim 1$ (i.e.\ large enough $|h|$).

\begin{figure}
    \centering
    \includegraphics[width=0.94\columnwidth]{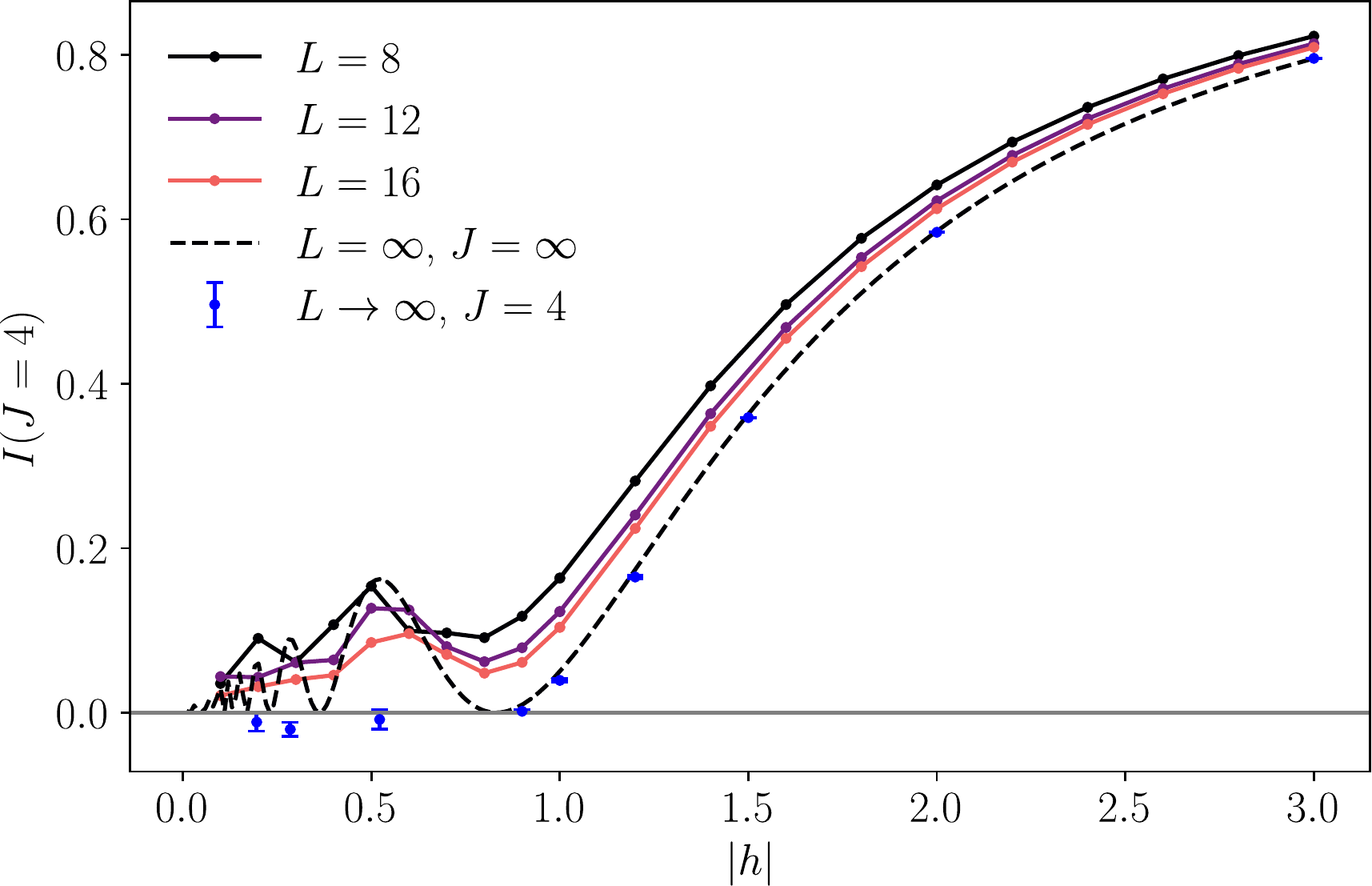}
    \caption{Imbalance~\eqref{eq:I-def} for $J=4$ (solid lines) and $J=\infty$ (dashed line) as a function of $h<0$, with $g=1$. The blue dots represent the imbalance extrapolated to $L \to \infty$ using the ansatz $I(L) = I_{\infty} + A/L$ (data for $L=6$,$10$,$14$ not shown in the plot). For $|h|<1$, the selected points correspond to the locations of the local maxima of $I(J=\infty)$, and are compatible with zero within one (at most two) standard deviations.}
    \label{fig:imbalance}
\end{figure}

To gain further insight on the perturbed dynamics of an interface, we consider an initial N\'eel state $\ket{\mathcal{N}} = \prod_{k \in \mathbb{Z}} \psi^{\dagger}_{2k}\ket{0}$---corresponding to a ``flat'' $45^\circ$-slope interface---evolving with $H^\prime_F$. We monitor the behavior of the time-averaged imbalance 
\begin{equation}
\label{eq:I-def}
    I =
    \lim_{L,T\to \infty}
    \frac 1 T
    \int_{0}^T dt\,
    \sum_{x=-L/2+1}^{L/2} \frac{1}{L} \bra{\mathcal{N}} m_x(t) m_x(0) \ket{\mathcal{N}} \, ;
\end{equation}
here $L$ is the length of the chain and $m_x(t)=2\psi_x^{\dagger}(t) \psi_x^{\phantom{\dagger}}(t)-1$. The quantity $I$ vanishes if the system thermalizes: $I\neq 0$ is thus an indicator of ergodicity breaking, which is often used in both numerical and experimental studies of many-body localization~\cite{Abanin2019Colloquium}. In the integrable limit $J\to\infty$ one 
finds $I(J=\infty)=J_0^2(2 g / h)$~\cite{Balducci2022Interface}. For finite $J$, we computed numerically $I(J)$ for finite chains $L<\infty$, extrapolating the data to $L\to\infty$. The result is shown in Fig.~\ref{fig:imbalance}: for $|h|/g \lesssim 1$ and finite $J$, the extrapolated  imbalance is compatible with zero (see inset); instead, for $|h|/g \gtrsim 1$, it is finite. While we are not able to conclude that the system is ergodic for $|h|/g \lesssim 1$ ($I=0$ is \emph{not} a sufficient criterion), for $|h|/g \gtrsim 1$ our results convincingly indicate the absence of thermalization. The threshold $|h|/g \simeq 1$ separating $I(J)>0$ from $I(J)=0$ appears to be weakly sensitive to the value of $J$ in the range $4 \le J < \infty$ (not shown). These results are fully consistent with the evidence of Stark {\it many-body} localization, analyzed in Refs.~\cite{Refael2019From,Schulz2019Stark}, for a setting very close to our emergent $1d$ fermionic Hamiltonian $H_{F}^\prime$.

\emph{Discussion and conclusions.~---}
In this work we introduced a promising approach for studying the dynamics of $2d$ quantum lattice models near a first-order phase transition, featuring competing vacuum states. In this context, large domains occupied by one vacuum separated from the competing vacuum by a ``smooth'' interface, naturally arise e.g.\ in prototypical nucleation scenarios~\cite{bray1994theory,onuki2002phase,Coleman1977False,coleman1988aspects,Rutkevich1999Decay}, and are easily prepared in modern experiments with quantum simulators~\cite{labuhn2016tunable,Guardado2018Probing,scholl2021quantum,ebadi2021quantum,Bluvstein2022Quantum}. For the ferromagnetic quantum Ising model we introduced a mapping from the interface onto a $1d$ fermionic chain. Hence, we demonstrated that the interface may exhibit robust non-ergodic behavior, underpinned by Stark many-body localization of the emergent fermionic excitations of the interface. Our findings thus establish a fascinating bridge between two apparently unrelated  dynamical phase transitions.

The approach presented here is expected to allow one to tackle even more generic questions concerning the dynamics of quantum interfaces in lattice models. For example, an intriguing issue, that we will address in a future work, is the ultimate ``evaporation'' of a bubble, driven by the exploration of disconnected configurations due to quantum fluctuations. Moreover, ``holographic" mappings of the kind introduced here may inspire experimental applications in which $2d$ degrees of freedom are used to engineer $1d$ Hamiltonians with interesting dynamical properties, or viceversa.

\begin{acknowledgments}

\emph{Acknowledgments.~---}
F.B.\ and C.V.\ would like to thank G.\ Giachetti, A.\ Santini and V.\ Vitale for discussions.
A.L. gratefully acknowledges stimulating discussions with A. Bastianello at the early stages of this project.
A.L.\ acknowledges support from the Swiss National Science Foundation.

\emph{Note added.~---}
While writing this work a preprint~\cite{Hart2022Hilbert} appeared discussing related issues. 

\end{acknowledgments}

\bibliography{references}

\newpage
\widetext
\begin{center}
\textbf{\large \emph{Supplemental Material for}\\ Localization and melting of interfaces in the two-dimensional quantum Ising model}
\end{center}
\setcounter{equation}{0}
\setcounter{figure}{0}
\setcounter{table}{0}
\makeatletter
\renewcommand{\theequation}{S\arabic{equation}}
\renewcommand{\thefigure}{S\arabic{figure}}

In this Supplemental Material we explain how Eqs.~\eqref{eq:hamFerm} and \eqref{eq:contLimit} of the main text can be derived on the basis of Eqs.~\eqref{eq:Ising2d_ham} and \eqref{eq:PXPEff}.

\section*{Derivation of Eq.~(3)}

According to the mapping between the interface of the $2d$ model and the $1d$ fermionic chain, each allowed spin flip (in the $\sigma^z$-basis) in $2d$---due to the term $ - g \sum_i \sigma^x_i $ in Eq.~\eqref{eq:Ising2d_ham}---corresponds to a nearest-neighbour fermion hop along the chain. This means  that the Hamiltonian governing the effective dynamics of the fermions along the chain contains an hopping term proportional to the spin-flip amplitude $-g$. The proportionality constant, that turns out to be $1$, is fixed by calculating a test matrix element of the Hamiltonian in $2d$ and by requiring that the corresponding matrix element on the fermionic chain gives the same result. Similarly, each spin flip in the configuration of the $2d$ system (see Eq.~\eqref{eq:Ising2d_ham}) implies a change in the average energy equal to $\pm 2h$, the sign depending on whether the spin flip is from down to up or vice-versa. Accordingly, each fermion hop implies a change in the total energy equal to $\pm 2h$, depending on whether it occurs to the left or to the right. This information leads directly to the Hamiltonian Eq.~\eqref{eq:hamFerm} of the main text. We remark that the Hamiltonian in Eq.~\eqref{eq:hamFerm} can be equivalently interpreted as ruling the dynamics of fermions or hard-core bosons, the equivalence residing in a standard Jordan-Wigner transformation.

\section*{Derivation of Eq.~(4)}

As discussed in the main text, to a given configuration of the fermionic chain specified by the occupation numbers $\{ n_i \}_i$  of the sites corresponds a unique (up to a constant) configuration of the interface of the $2d$ spin system in the $\sigma^z$-basis which runs at a distance $\sum_{j\le i} (2 n_j - 1) + \mbox{const}$  from the reference line, as in Fig.~\hyperref[fig:PXPHam]{\ref{fig:PXPHam}d} of the main text. Accordingly, the average position of the fluctuating interface in the $2d$ model can be obtained from the expectation value $\langle \mu(x) \rangle$ of the corresponding operator $\mu(x)$ defined on the chain and introduced in the main text, which involves the average fermion number $\langle \psi^\dagger_x(t)\psi_x(t)\rangle$. In turn, the evolution of this density can be easily obtained once the Hamiltonian of the chain is diagonalized on the basis of the eigenstates $\phi_x^{(y)}$, finding 
\begin{equation}
    \langle \psi^\dagger_x(t)\psi_x(t)\rangle = \sum_{y,z} e^{i(z-y)ht + i \pi(z-y)/2} J_{x-y}(\omega_t) J_{x-z}(\omega_t) \langle \psi^\dagger_y(0)\psi_z(0)\rangle,
\end{equation}
where we introduced $\omega_t := 2 g |\sin(ht)/h|$. Specifying this evolution to the case of the initial state $|\Psi_0\rangle$ and after taking the continuum limit, we find an average fermion density (for the case $h=0$, see Ref.~\cite{Antal1999Transport})
\begin{equation}
    n(x,t) =
    \begin{cases}
        0       &\mbox{for}\quad  x < - \omega_t,\\
        \frac{1}{2} + \frac{1}{\pi}\arcsin(x/\omega_t)
        &\mbox{for}\quad |x| \leq \omega_t, \\
        1       &\mbox{for}\quad x > \omega_t.
    \end{cases}
\end{equation}
An integration over $x$ (which is the limit of the above sum on the continuum) leads to the expression for the average height operator in Eq.~\eqref{eq:contLimit}.


\end{document}